# An experimental assessment of displacement fluctuations in a 2D granular material subjected to shear

**Vincent Richefeu (1), Gaël Combe (1), Gioacchino Viggiani (1)**

Email Address of Submitting Author:

gael.combe@3sr-grenoble.fr

Postal Address of Submitting Author:

Laboratoire 3SR
Domaine Universitaire
BP53
38041 Grenoble
France

**Author Affiliations:**

(1) Grenoble-INP/UJF-Grenoble 1/CNRS UMR 5521, Laboratoire 3SR, Grenoble, France

**Abstract:** In a granular material, a macroscopically homogeneous deformation does not correspond to a homogeneous displacement field when looking at the individual grains. The deviation of a grain displacement from the value dictated by the continuum field (referred to as *fluctuation*) is likely to hold valuable information about the characteristic length(s) involved in grains' rearrangement, which is the principal mechanism of irreversible deformation for granular materials. This paper shows a selection of results from a series of shear tests on a 2D analogue granular material. We have followed the route opened by the pioneering work of Radjaï and Roux (2002), and used the same framework to analyze our *experimental* data on displacement fluctuations. Digital Image Correlation has been used to measure and characterize the displacement fluctuations. The analysis of their spatial organization reveals the emergence of a minimum length scale that is in the order of 10 times the mean particle size.

# Introduction

In a granular material, a macroscopically homogeneous deformation does not correspond to a homogeneous displacement field when looking at the individual grains. Due to geometrical constraints at the grain scale (mutual exclusion of grain volumes), grains are not able to displace as continuum mechanics dictates they should. This can be pictured by imagining an individual in a crowd of people who all wish to go to the same place: although the long-term displacement of each individual is equal to the displacement of the crowd, the steps of each individual are erratic. Classical continuum approaches disregard this feature, which is apparent only at strains that are 'small' (*i.e.*, for displacements that are small with respect to the size of rigid grains, or the size of contact indentation for deformable grains). However, the deviation of a grain's displacement from the value dictated by the continuum field (referred to as *fluctuation*) is likely to hold valuable information about the characteristic length(s) involved in grains' rearrangement, which is the principal mechanism of irreversible deformation in granular materials.

Displacement fluctuations, often spatially organized in the form of vortices, have been observed in (quasistatic) experiments only by Misra & Jiang (1997). Several numerical studies using Discrete Elements (DEM) report similar observations (*e.g.*, Williams & Rege, 1997; Kuhn, 1999; Combe & Roux, 2003; Tordesillas *et al.,* 2008; Rechenmacher *et al*., 2011;). Radjaï & Roux (2002) investigated such fluctuations by making use of statistical analysis that is typically used in fluid dynamics. The main finding of their study is that displacement fluctuations in granular materials show scaling features having striking analogies with fluid turbulence. Inspired by the approach of Radjaï & Roux (2002), the present study analyzes fluctuations *measured from experiments* on a 2D analogue granular material subjected to (quasi-static) shear.

# Shear Test

A number of strain controlled shear tests have been performed in the 1γ2ε apparatus, which is essentially a plane stress version of the directional shear cell developed for testing soils (see Joer *et al.* 1992 and Calvetti *et al.* 1997 for details). Results from only one such test are discussed herein, although the results are consistent throughout the entire campaign. In this experiment, the assembly of 2D grains is slowly deformed in simple shear at constant vertical stress $\sigma_n$ = 50 kPa (Figure 1). Note that the contact properties among the (wooden) grains compare with the DEM-imposed properties used by Radjai and Roux (2002). The vertical sides of the enclosing frame are tilted up to $\gamma$ = 0.26 (15°) while the length of the horizontal sides is kept constant. The shear strain rate is $8.2\times10^{-5}$ $s^{-1}$, which is small enough to ensure a quasi-static regime according to the inertial number criterion suggested by Combe & Roux (2003). A video of the test can be found at http://youtu.be/B4Dfesn5vhs .

During the test, normal ($\sigma_n$) and tangential ($\sigma_t$) stresses are measured at the boundary of the sample (here on the top rigid wall, as illustrated by two vectors on Figure 2a). The global stress-strain response is typical of a dense 2D granular material, with a peak friction angle of about 26° at $\gamma \approx 0.06$ and a dilatant behaviour throughout (Figure 2c). A digital camera was used to acquire 24.5 MPixels images every 5 seconds (*i.e.*, a shear strain increase of $\Delta\gamma = 4\times10^{-4}$, called *strain window* in the sequel) all along the test. As an example, Figure 2b shows the displacements of all grains measured by DIC from 0 to 0.12 shear strain $\gamma$.

# Particle tracking

A novel code has been developed to process the digital images and measure (with sub-pixel resolution) the in-plane displacement and rotation of each individual grain from one image to another. This code, called *Tracker*, uses a discrete, grain-scale version of Digital Image Correlation (DIC) that is well suited for tracking rigid particles. A grain is tracked from the first image to any other image by (1) choosing a region (smaller than the grain) around the center of the grain and then (2) finding the rigid body motion (translation and rotation) of this region that best maps the initial image onto the current image. A remarkable feature of *Tracker* is the use of a specific strategy that allows tracking *all* particles from one image to another, without 'losing' any of them (which is a typical problem when tracking assemblies of discrete particles over many images). In essence, this is achieved by a two-step procedure, where in case of problematic tracking of a particle, the size of the search zone is increased in an adaptive manner, *i.e.*, taking into account the results of tracking in the neighborhood of the particle.

Since this study focuses on fluctuations, which can be very small (relative to the pixel size, which is equal to 0.11 mm in the images analyzed herein), the accuracy of the measurement technique is a crucial issue. Assessing the accuracy of *Tracker* with real images of a deforming assembly of grains is complex because it involves the quality of the images themselves. A global measure of accuracy (integrating different sources of error) can be obtained by tracking four points (placed on 200 grains) that define two initially perpendicular virtual line-segments (see inset in Figure 3a). If the grains are perfectly rigid, then the lengths of the two segments and the angle between them should not change during the deformation of the specimen. Figures 3a and 3b show the average and standard deviation of the measured variations of length ($\Delta S$) and angle ($\Delta\alpha$), respectively, throughout the test – obtained by comparing each image to the initial one. It can be seen that $\Delta S = 0.01\pm0.05$ pixel (or $1.1\times10^{-3}\pm5.5\times10^{-3}$ mm) and $\Delta\alpha = 0.01\pm0.06$ degrees, which defines the confidence that we can have in the measured fluctuations.

## Assessment of fluctuations

To assess the fluctuations of displacement (angular rotations of grains are also assessed but not discussed in the paper) in the course of deformation, we consider two possible displacements of each grain during a strain window $\Delta\gamma$ (always positive). The first is the actual displacement vector $\delta\boldsymbol{r}(\gamma,\Delta\gamma)$, which depends both on the size of the strain window $\Delta\gamma$ and the level of shear strain $\gamma$ at the beginning of the strain window. The second displacement vector $\delta\boldsymbol{r}^*(\gamma,\Delta\gamma)$ is the displacement dictated by a homogeneous (*affine*) continuum strain field, *i.e.*, the displacement that the grain's center *would* have if it moved as a material point in a continuum. The fluctuation of the displacement is defined as the difference between these two displacements vectors:

$$\mathbf{u}(\gamma,\Delta\gamma) = \delta\mathbf{r}(\gamma,\Delta\gamma) - \delta\mathbf{r}^*(\gamma,\Delta\gamma) \qquad (1)$$

Displacement fluctuations can be conveniently normalized by dividing the vector $\boldsymbol{u}(\gamma,\Delta\gamma)$ by the product $\Delta\gamma\,\langle d\rangle$ (where $\langle d\rangle$ is the mean diameter of the grains), which can be interpreted as the average displacement of the grains in the strain window $\Delta\gamma$. This normalized fluctuation:

$$\mathbf{V}(\gamma,\Delta\gamma) = \frac{\mathbf{u}(\gamma,\Delta\gamma)/\langle d\rangle}{\Delta\gamma} \qquad (2)$$

can also be interpreted as a *local* (microscopic) strain fluctuation which is in turn divided by the size of the *global* (macroscopic) strain window $\Delta\gamma$.

Figure 4a shows the mean magnitude of fluctuations, $\langle u\rangle$, and the mean magnitude of normalized fluctuations, $\langle V\rangle$, as a function of $\Delta\gamma$ (taken from $\gamma = 0$). The mean displacement fluctuation increases monotonically throughout the test, from 0.066 (for $\Delta\gamma = 10^{-3}$) to 6 mm (for $\Delta\gamma = 0.25$), this final value corresponding to about 10% of the average displacement of the grains between the beginning and the end of the shear test. Note that the smallest displacement fluctuation is well above the accuracy of *Tracker*. As far as the mean magnitude of normalized fluctuations $\langle V\rangle$ is concerned, a decrease of $\langle V\rangle$ from 3 to 2 is observed; this implies that the average fluctuation of local shear strain $\langle u\rangle/\langle d\rangle$ is two to three times larger than the global strain window $\Delta\gamma$. Figure 4b suggests that there is a direct link between the changes in $\langle V\rangle$ and the volumetric strain of the granular assembly; this is consistent with the very origin of fluctuations, which are due to geometrical constraints at the grain scale.

### Selected results

Figure 5 shows the probability density function (*pdf*) of normalized fluctuations for two different strain window $\Delta\gamma$. Whatever the value of $\Delta\gamma$, the space-average of this measure of fluctuations is zero, which is to be expected for a global homogeneous deformation. A key observation is that the *pdf* curve exhibits a wider range of fluctuations with decreasing $\Delta\gamma$. The inset in Figure 5 shows the kurtosis of the *pdf* as a function of $\Delta\gamma$. Kurtosis is seen to tend to zero asymptotically with increasing size of $\Delta\gamma$. A zero value of kurtosis might indicate that the *pdf* tends to become Gaussian for $\Delta\gamma > 0.05$. Very similar features and trends have been observed (on DEM simulations) by Radjaï & Roux (2002), who noted that they have striking similarities with what is observed in turbulent flow of fluids (although *no* dynamics are associated with these fluctuations). For a deeper description of the physics-oriented tools used in the field of fluids turbulence, readers are referred to the book of Frisch (1995).

In order to understand fluctuations, the analysis of their statistics must be supplemented by the study of their spatial distribution. The vectors in Figures 6a-e are the normalized fluctuations at different shear strain levels for $\Delta\gamma = 10^{-3}$. At every shear strain level, fluctuations are clearly organized in space, with vortex-like structures which are reminiscent of the patterns of turbulence in fluids – with the important difference that in fluids, vortices are exhibited by the displacements/velocities themselves, while here they are present in the fluctuations. Contrary to what happens in fluid dynamics, the largest fluctuations are on the periphery rather than in the core of the vortex. The space-correlation of these fluctuations (referred to as *fluctuation loops* hereafter) is investigated by computing the auto-correlation coefficient for each pair of fluctuating grains separated by a given distance (correlograms). Figure 6f shows the correlation coefficient (*Pearson* coefficient) as a function of such distance (expressed as a multiple of mean particle diameters $\langle d \rangle$) for each of the five fluctuation maps in Figures 6a-e. The five curves are different from each other, but for all of them correlation decreases with increasing distance and a strong correlation is found for distances less than 10 particle diameters. This length roughly corresponds to the minimum radius of the fluctuation loops. A pseudo-period of about 20 diameters is observed for two out of the five curves (pink and black in Figure 6f), which is another signature of the loops. This pseudo-periodicity might indicate the existence of a *cascade* of loop-sizes rather than a unique size – the smallest size being about 10 diameters, and the largest in the order of the sample size. This interpretation is confirmed by performing a Fourier transform of the fluctuation signal as a function of space as suggested by Radjaï & Roux (2002). As an example, Figure 7 shows the power spectrum of the horizontal component of $\boldsymbol{V}$ along the horizontal direction for $\Delta\gamma = 10^{-1}$ and $\Delta\gamma = 10^{-3}$. For both values of $\Delta\gamma$, power $E$ decays with increasing (nondimensional) frequency $k$, exhibiting a power-law $\sim k^{-5/3}$, which suggests that the observed spatial patterns are self-similar structures (see, e.g., Feder 1988). The parameter of similitude 5/3 is here unexpectedly the same as the one observed in fluid turbulence; note that Radjaï and Roux (2002) obtained a value of 5/2 from their simulations. Interestingly, the power-law only applies to frequencies lower than $10^{-1}$, *i.e.*, to distances equal to or greater than 10 mean particle diameters – which is consistent with what observed in Figure 6f.

It is of interest to also investigate whether the patterns of fluctuation change when changing the size of $\Delta\gamma$ (which can be thought of as an *observation window*). The vectors in Figures 8a-e are the normalized fluctuations of displacement for different sizes of the observation window for $\gamma = 4\times10^{-2}$. The fluctuation loops are increasingly similar to each other with increasing $\Delta\gamma$. This is also confirmed by the correlograms in Figure 8f, which become closer and closer to each other with increasing $\Delta\gamma$. A quantitative assessment of the observed similarity of these fluctuation loops was also carried out, by using the two-dimensional Kolmogorov-Smirnov test (which is a standard test in statistics, see Press *et* *al.* 1986). Results of this analysis, not shown herein in the interest of space, indicate that the larger the observation window, the longer the patterns 'survive'. An illustration of this observation can be seen on the two videos available at: http://youtu.be/q_Q0vIemxxc and http://youtu.be/s6k8q6x8Zx4 .

However, no single value of the observation window size has been found that would be *characteristic* of the fluctuation patterns, *i.e.*, there seems to be no evidence of a single strain-scale.

## Conclusions

An original DIC technique has been developed which is capable of accurately measuring the fluctuations of particle displacements in a 2D granular material subjected to quasi-static shear. A detailed study of such fluctuations has revealed that they are organized in space and show clear vortex-like patterns, reminiscent of turbulence in fluid dynamics. While these fluctuation loops have been often observed in DEM numerical simulations, on the experimental front they have been reported (to the authors' best knowledge) only in the study by Misra & Jiang (1997). The present study confirms these early findings and brings into the picture the important idea that fluctuation loops have a characteristic minimum radius – equal to 10 mean particle diameters for the material tested.

Displacement fluctuations in granular materials are a direct manifestation of grain rearrangement, therefore they can be thought of as the basic mechanism of irreversible deformation – they play a similar role as dislocations in crystals. The link between these fluctuations (and their spatial organization) and the deformation of granular materials at the macro scale (including dilatancy, strain localization, critical state) remains still to be investigated.

## Acknowledgements

This study was inspired by the work of Farhang Radjaï, and actually started with the master project of Alessandro Tengattini; a number of interesting discussions took place – at different stages during the study – with Itai Einav, Steve Hall, Dimitrios Kolymbas and Edward Andò. All these colleagues are gratefully acknowledged.

## Figures

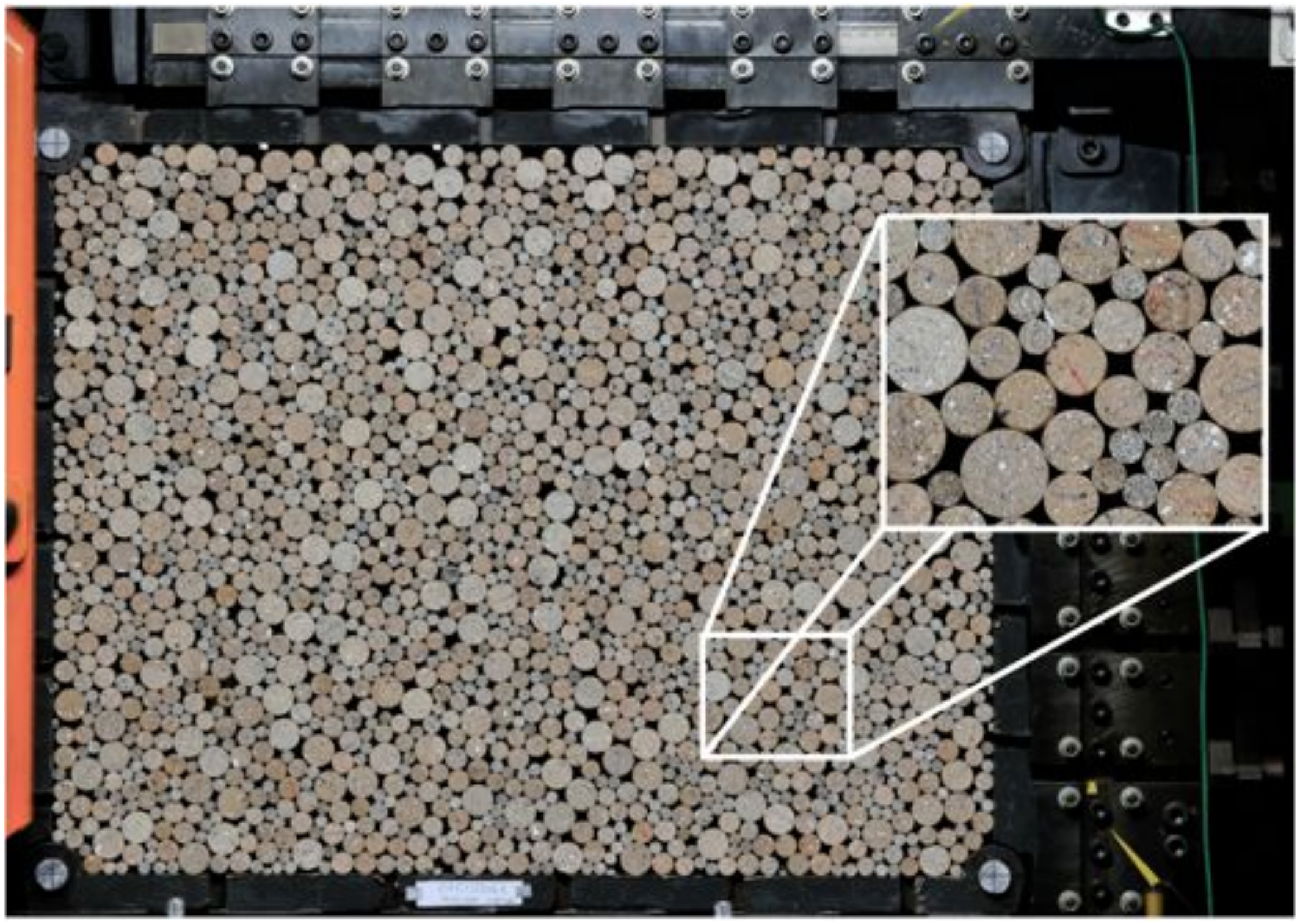

Fig. 1.
*Photograph of a specimen made up of about 2000 2D 'grains' (wooden cylinders 6 cm long) having four different diameters, from 8 to 20 mm, under isotropic loading. The granular packing is enclosed by a rigid rectangular frame of 0.56 m x 0.47 m. A speckle of black and white points is painted on each cylinder (see inset-figure) to allow the measurement of particle kinematics by means of Digital Image Correlation.*

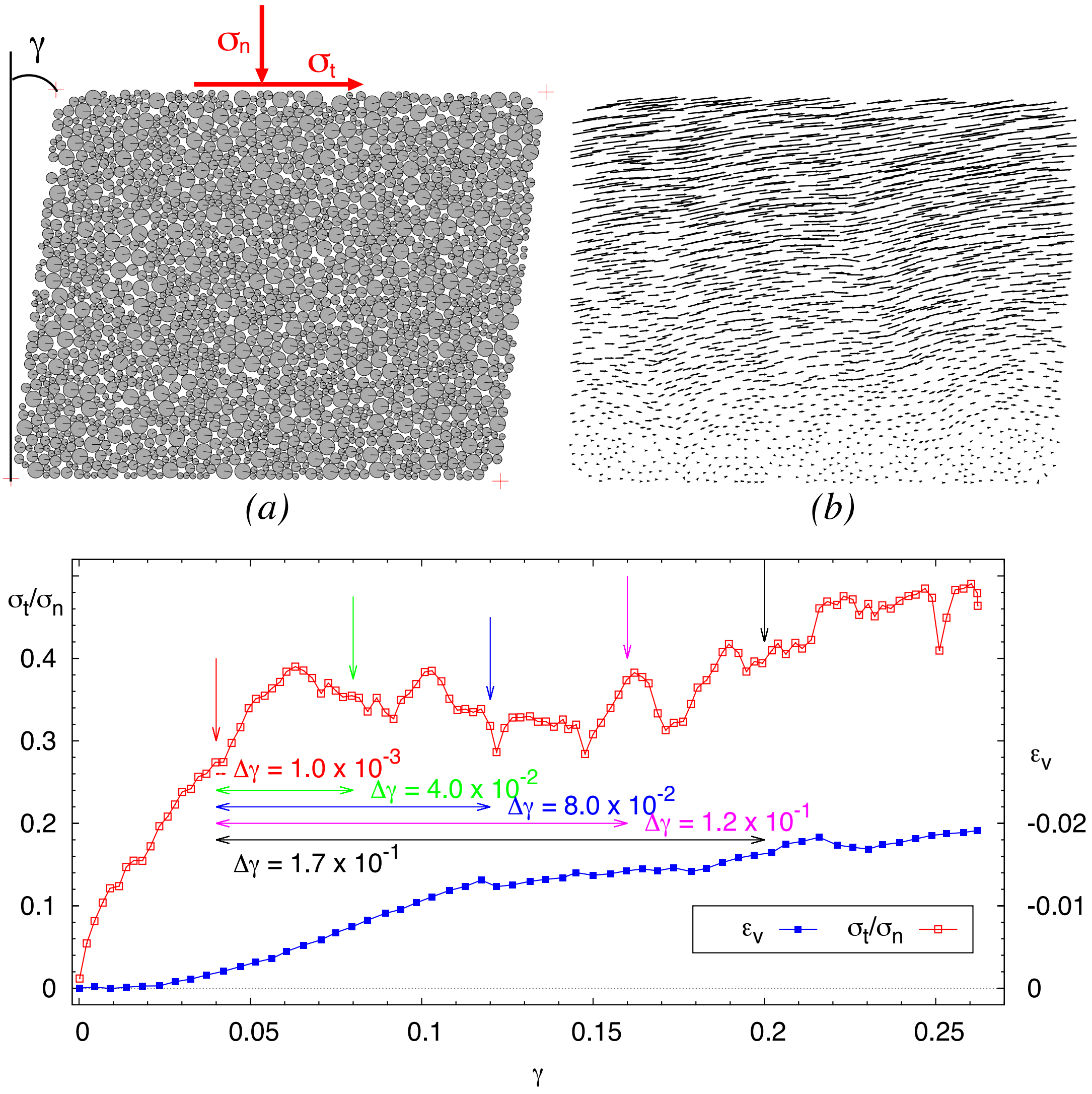

(c)

Fig. 2.

*(a) DIC-based image of granular assembly at $\gamma$ = 0.12; (b) displacements of individual grains from $\gamma$ = 0 to $\gamma$ = 0.12 (maximum displacement is approximately 60 mm); (c) stress ratio and volumetric strain ($\varepsilon_v$ > 0 for contraction) as a function of shear strain.*

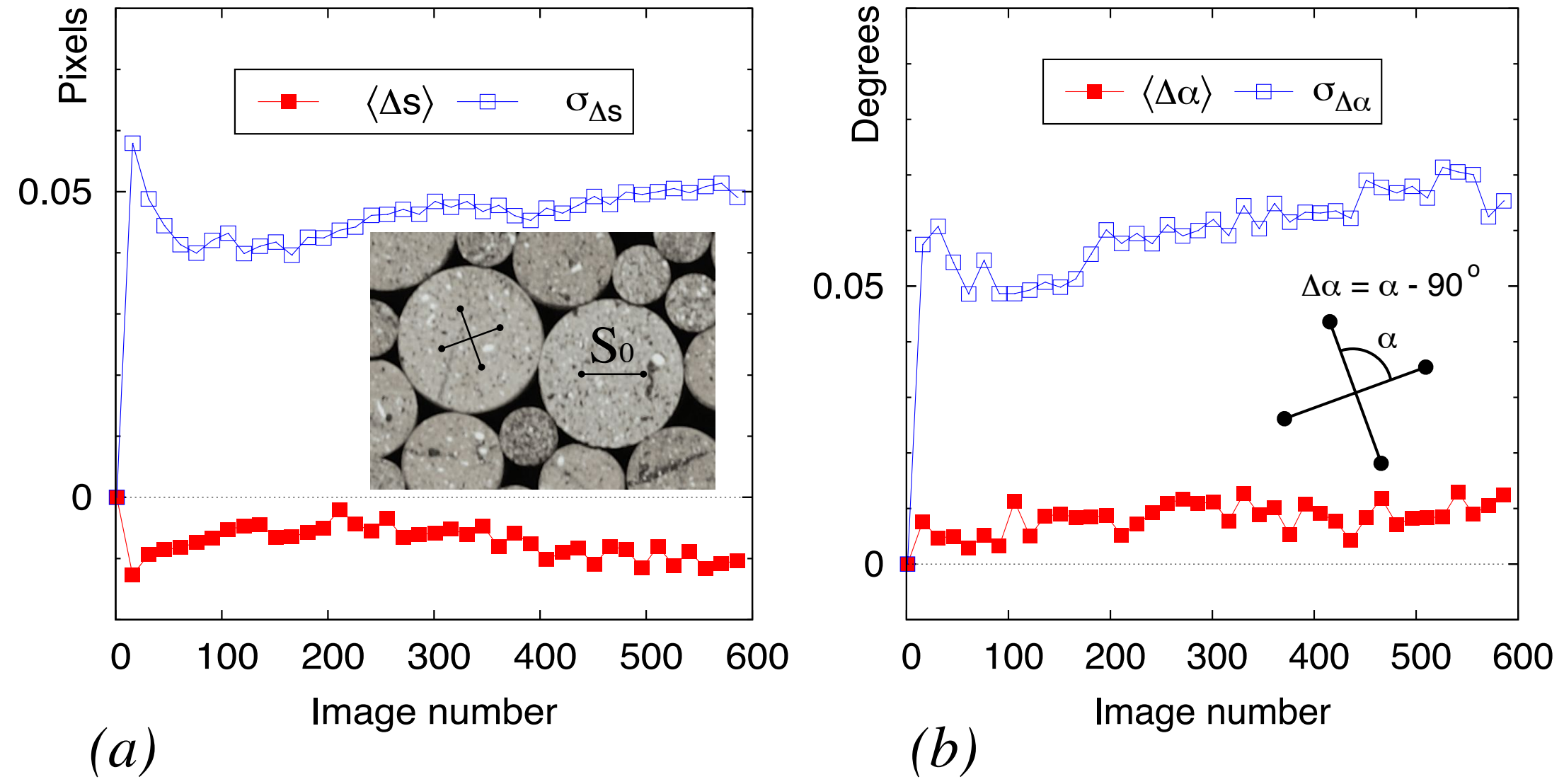

Fig. 3.
*Accuracy of particle tracking: (a) average (red) and standard deviation (blue) of measured changes of length ΔS of 434 segments (initial length 60 pixel; (b) average (red) and standard deviation (blue) of measured changes of angle* $\Delta\alpha$ *of 217 pairs of (initially perpendicular) segments.*

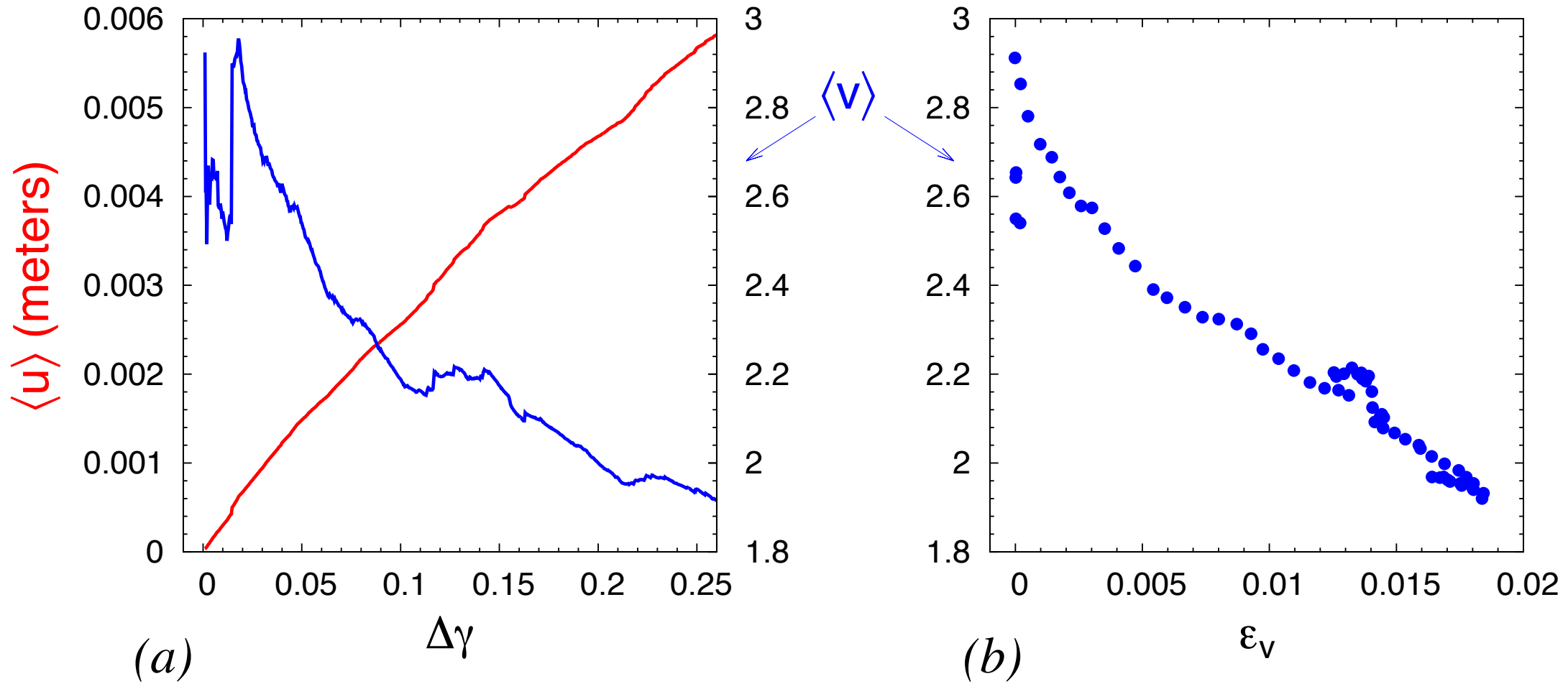

Fig. 4.
*(a) evolution of the mean values of the norm of fluctuations, ⟨u⟩ and normalized fluctuations, ⟨V⟩ as a function of* $\Delta\gamma$ *(at* $\gamma = 0$*); (b) evolution of ⟨V⟩ with the volumetric strain of the sample.*

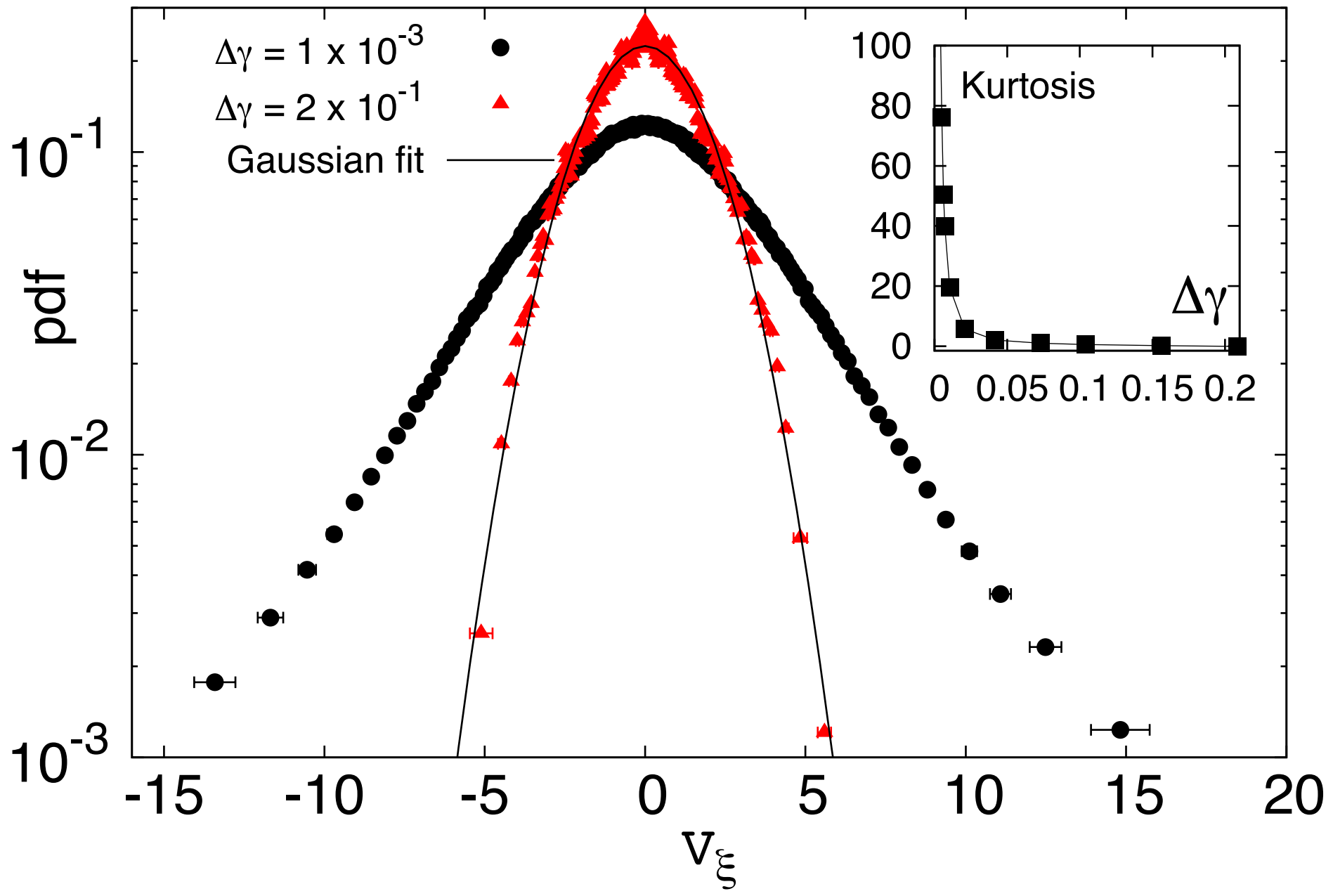

Fig. 5.

*Probability density function (pdf) of normalized fluctuations $V_\xi$ projected on the major principal direction $\xi$ of strain for two different strain windows, $\Delta\gamma = 10^{-3}$ and $\Delta\gamma = 2\times10^{-1}$. The major principal direction is given by $\xi = \tan^{-1}[\gamma/(\varepsilon_x - \varepsilon_y)]$. The inset shows the kurtosis of the pdf as a function of $\Delta\gamma$.*

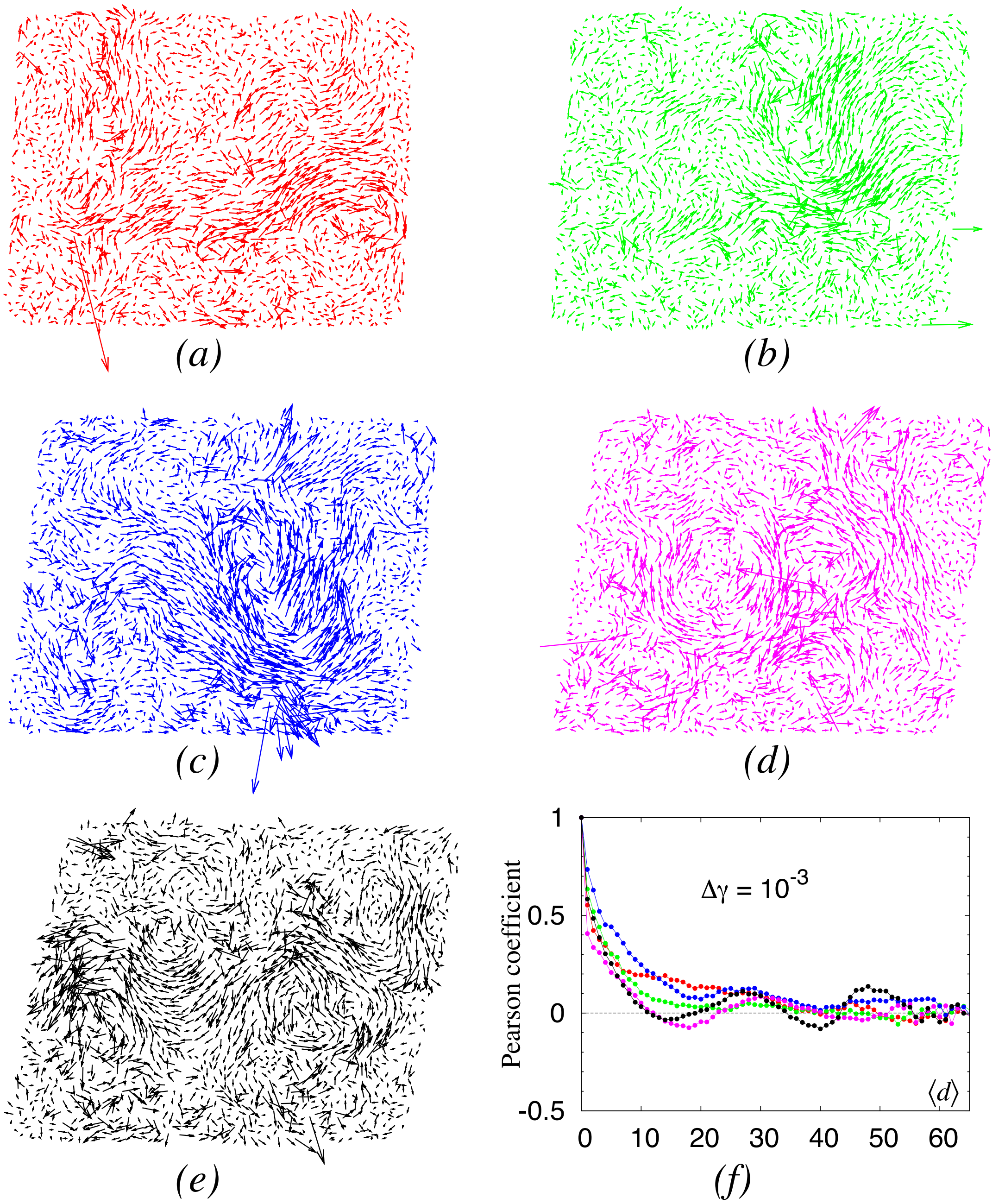

Fig. 6.
(a) − (e): *maps of normalised fluctuations* **V** *for one constant strain window $\Delta\gamma$ = $10^{-3}$ at different values of $\gamma$ (corresponding to the colored vertical arrows in Figure 2c). (f): Spatial correlograms computed for each fluctuation map (Pearson coefficient as a function of mean grain diameter ⟨d⟩).*

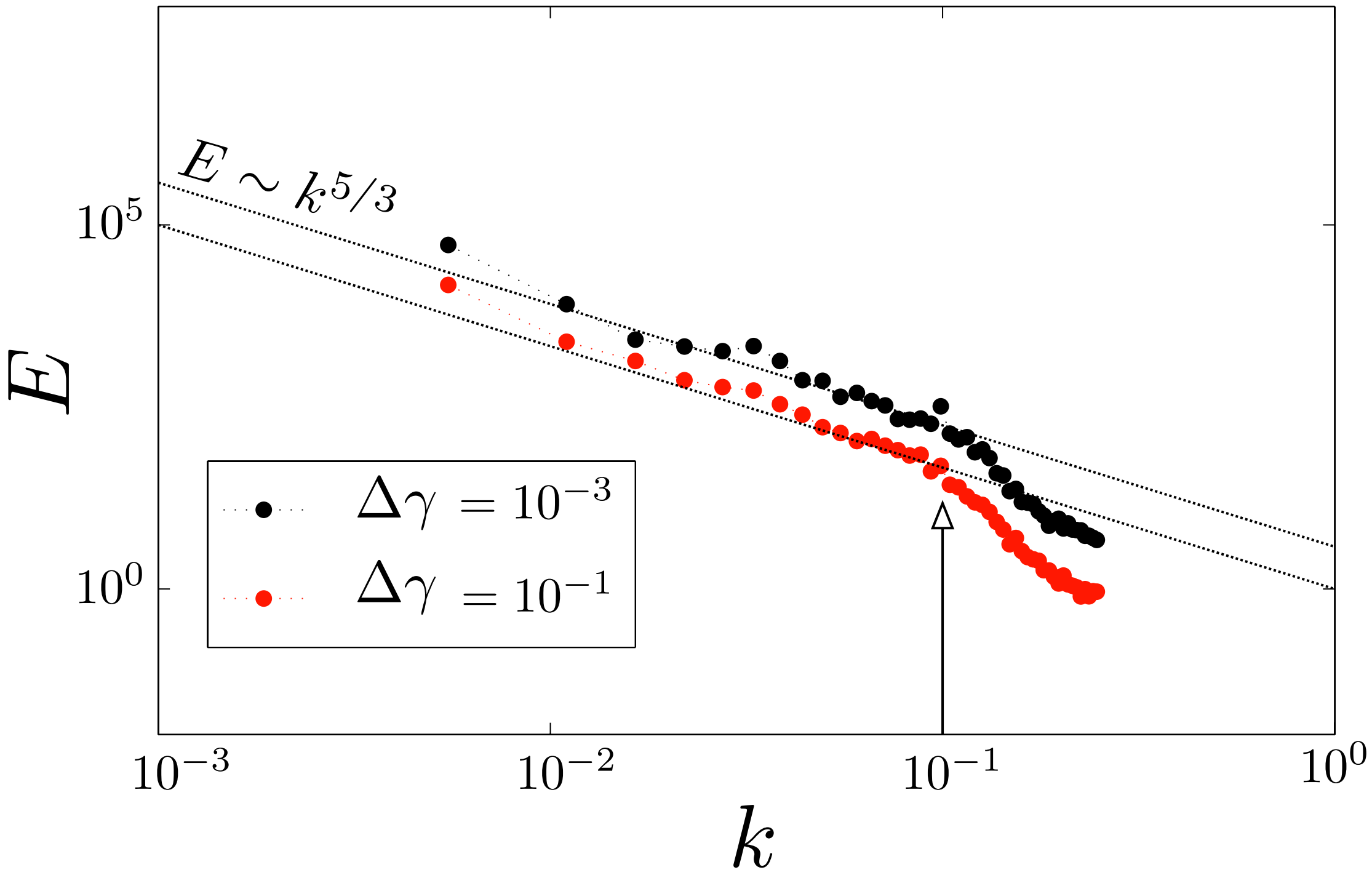

Fig. 7.
*Power spectrum of the horizontal component of V along the horizontal direction for $\Delta\gamma = 10^{-3}$ and $\Delta\gamma = 10^{-1}$ (frequency* k *is the inverse of a length expressed as a multiple of mean particle diameters).*

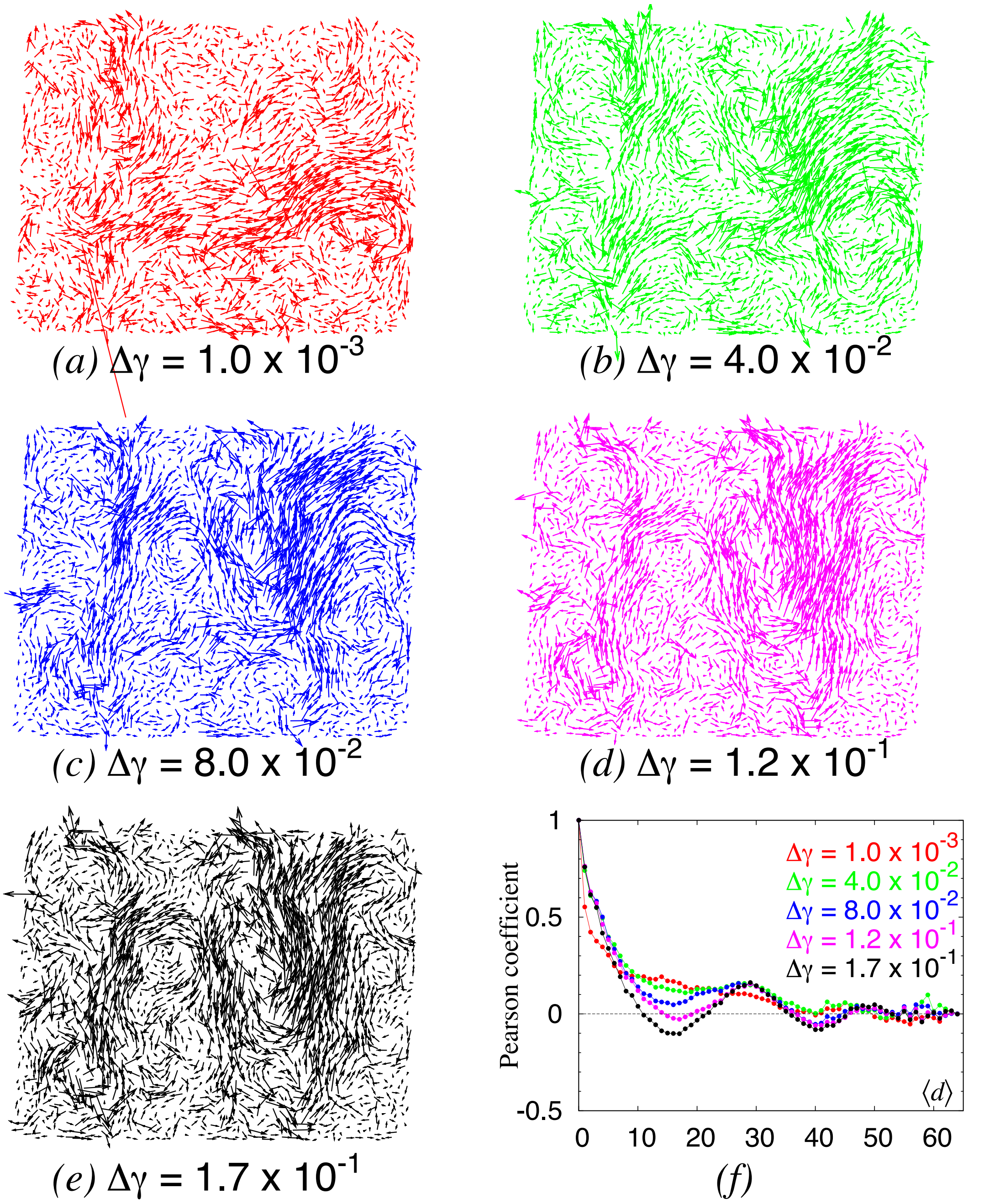

Fig. 8.

*(a) − (e): maps of normalised fluctuations* ***V*** *for different strain window $\Delta\gamma$ from $\gamma = 0.04$ to $\gamma = 0.04 + \Delta\gamma$ (corresponding to the colored horizontal arrows in Figure 2c); (f): Spatial correlograms computed for each fluctuation map (Pearson coefficient as a function of mean grain diameter $\langle d \rangle$).*